\documentclass[aps,pra,superscriptaddress,twocolumn,longbibliography,notitlepage]{revtex4-2}
\usepackage{mathrsfs}
\usepackage{amsfonts}
\usepackage{amssymb}
\usepackage{amsmath}
\usepackage{graphicx}
\graphicspath{{figures/}{./}} 
\usepackage{sidecap}
\usepackage{color}
\usepackage[colorlinks=true, urlcolor=blue, citecolor=blue,linkcolor=blue,citebordercolor={1 0 0},linkbordercolor={0 0 1}]{hyperref}
\usepackage{subeqnarray}
\usepackage{verbatim}
\usepackage{euscript} 

\usepackage{hyperref}

\usepackage{bibunits}

\usepackage{array}
\usepackage{multirow}

\usepackage{blindtext}

\usepackage{graphicx}
\usepackage{dcolumn}
\usepackage{bm}

\renewcommand{\eqref}[1]{Eq.~(\ref{#1})} 
\newcommand{\figref}[1]{Fig.~\ref{#1}} 

\usepackage{graphicx}
\usepackage{physics}
\usepackage{amsmath}
\usepackage{tikz}
\usepackage{mathdots}
\usepackage{yhmath}
\usepackage{cancel}
\usepackage{color}
\usepackage{siunitx}
\usepackage{array}
\usepackage{multirow}
\usepackage{amssymb}
\usepackage{gensymb}
\usepackage{tabularx}
\usepackage{extarrows}
\usepackage{booktabs}
\usetikzlibrary{fadings}
\usetikzlibrary{patterns}
\usetikzlibrary{shadows.blur}
\usetikzlibrary{shapes}

\DeclareMathOperator*{\Motimes}{\text{\raisebox{0.25ex}{\scalebox{0.8}{$\bigotimes$}}}}

\tikzset{every picture/.style={line width=0.75pt}} 

\usepackage{float}

\newcommand{\rle}{Research Laboratory of Electronics, Massachusetts Institute of Technology, Cambridge, MA 02139, USA}
\newcommand{\mitphysics}{Department of Physics, Massachusetts Institute of Technology, Cambridge, MA 02139, USA}
\newcommand{\miteecs}{Department of Electrical Engineering and Computer Science, Massachusetts Institute of Technology, Cambridge, MA 02139, USA}
\newcommand{\harvardphysics}{Department of Physics, Harvard University, Cambridge, MA 02138, USA}
\newcommand{\mitll}{Lincoln Laboratory, Massachusetts Institute of Technology, Lexington, MA 02421-6426, USA}

\begin{document}

\preprint{APS/123-QED}

\title{Distinguishing multi-spin interactions from lower-order effects}

\author{Thomas R. Bergamaschi}
\email{thomasbe@mit.edu}
\affiliation{\mitphysics}

\author{Tim Menke}
\email{timmenke@mit.edu}
\affiliation{\mitphysics}
\affiliation{\rle}
\affiliation{\harvardphysics}

\author{William P. Banner}
\affiliation{\miteecs}

\author{Agustin Di Paolo}
\affiliation{\rle}

\author{Steven J. Weber}
\affiliation{\mitll}

\author{Cyrus F. Hirjibehedin}
\affiliation{\mitll}

\author{Andrew J. Kerman}
\affiliation{\mitll}

\author{William D. Oliver}
\email{william.oliver@mit.edu}
\affiliation{\mitphysics}
\affiliation{\rle}
\affiliation{\miteecs}
\affiliation{\mitll}

\date{\today}

\begin{abstract}
    Multi-spin interactions can be engineered with artificial quantum spins.
    However, it is challenging to verify such interactions experimentally.
    Here we describe two methods to characterize the $n$-local coupling of $n$ spins.
    First, we analyze the variation of the transition energy of the static system as a function of local spin fields.
    Standard measurement techniques are employed to distinguish $n$-local interactions between up to five spins from lower-order contributions in the presence of noise and spurious fields and couplings.
    Second, we show a detection technique that relies on time dependent driving of the coupling term.
    Generalizations to larger system sizes are analyzed for both static and dynamic detection methods, and we find that the dynamic method is asymptotically optimal when increasing the system size.
    The proposed methods enable robust exploration of multi-spin interactions across a broad range of both coupling strengths and qubit modalities.
\end{abstract}

\maketitle

\textit{Introduction}.---Although interactions that depend on more than two spins are usually not present in natural quantum spin systems, they are a key resource for analog quantum simulation and quantum computation applications.
For example, multi-spin interactions arise in the mapping of fermions to artificial spin systems in quantum algorithms for quantum chemistry \cite{Seeley2012, Babbush2014, OMalley2016} or in effective spin models of  cuprate superconductor Hamiltonians \cite{Majumdar2012}. In addition, multi-spin interactions play a crucial role in error suppression schemes for quantum annealers \cite{Bacon2006, Kerman2018}, in adiabatic topological quantum computation \cite{Cesare2015}, as well as in a variety of combinatorial optimization problems \cite{Marto_k_2004,Santoro_2006,johnson2011quantum}.
Such multi-spin interactions are called $n$-local when there are $n$ spins partaking in a single many-body coupling mechanism.

While many $n$-local coupling proposals use perturbative Hamiltonian gadgets with ancilla qubits to obtain the effective interaction \cite{kempe2006complexity,jordan2008perturbative,leib2016transmon,Chancellor_2017}, a variety of non-gadgetized 3- and 4-local interaction schemes have been proposed \cite{Mezzacapo_2014,Hafezi_2014,Schondorf_2019,melanson2019tunable,menke2019automated,liu2020synthesizing}, some of which are scalable to higher-order locality. 
While the implementation of such interactions appears to be of broad interest, the identification of multi-spin interactions when these are relatively weak and appear in combination with interactions of other orders that are of comparable strength or larger remains an open question.
In addition, some coupling mechanisms can cause further undesired couplings of lower order \cite{Schondorf_2019,melanson2019tunable,menke2019automated}, requiring detection methods that reliably determine the coupling despite such spurious terms.

In this work, we propose error-resilient methods for the characterization of multi-spin interactions. The studied methods can be split into two classes: one relying on the static properties of an $n$-local Hamiltonian, and the other exploiting time dynamic effects when driving such a Hamiltonian. For the static detection technique, we rely only on spectroscopic measurements rather than temporal dynamics.
Naturally, this method is directly relevant for experiments with artificial spin systems that exhibit limited coherence.
Using Hamiltonian model fits, which are commonly employed to characterize artificial spin systems \cite{Weber_2017, Harris_2007, Ozfidan_2020}, we find numerical evidence that the $n$-local interaction adds a unique signature to the behavior of the transition energy of the system when local spin fields are varied.
We study the robustness of this technique by analyzing the performance of the method in the presence of emulated experimental noise, which dampens the size of the $n$-local coupling signature.This static method is contrasted with a time dependent detection technique that requires a stronger set of assumptions on the experimental system control.
In this case, the capability of detecting $n$-local interactions scales more favorably with system size, and we find that it is an asymptotically optimal detection technique.
However, it is subject to stringent coherence time requirements.
By contrasting detection method classes with high and low experimental control requirements, we identify their respective merits for practical quantum device characterization.

\begin{figure}[htb]
    \centering
\begin{tikzpicture}[x=0.75pt,y=0.75pt,yscale=-0.5,xscale=0.5]

\draw  [fill={rgb, 255:red, 155; green, 155; blue, 155 }  ,fill opacity=0.43 ] (205.84,73.09) .. controls (205.84,47.13) and (226.89,26.09) .. (252.84,26.09) -- (434.87,26.09) .. controls (460.82,26.09) and (481.87,47.13) .. (481.87,73.09) -- (481.87,214.09) .. controls (481.87,240.05) and (460.82,261.09) .. (434.87,261.09) -- (252.84,261.09) .. controls (226.89,261.09) and (205.84,240.05) .. (205.84,214.09) -- cycle ;
\draw [color={rgb, 255:red, 208; green, 2; blue, 27 }  ,draw opacity=1 ][line width=1.5]    (615.01,457.91) .. controls (587.39,383.12) and (491.86,338.15) .. (451.27,322.99) ;
\draw [shift={(447.67,321.67)}, rotate = 379.75] [fill={rgb, 255:red, 208; green, 2; blue, 27 }  ,fill opacity=1 ][line width=0.08]  [draw opacity=0] (11.61,-5.58) -- (0,0) -- (11.61,5.58) -- cycle    ;
\draw [shift={(616.67,462.67)}, rotate = 252] [fill={rgb, 255:red, 208; green, 2; blue, 27 }  ,fill opacity=1 ][line width=0.08]  [draw opacity=0] (11.61,-5.58) -- (0,0) -- (11.61,5.58) -- cycle    ;
\draw [color={rgb, 255:red, 208; green, 2; blue, 27 }  ,draw opacity=1 ][line width=1.5]    (445.68,450.4) .. controls (468.34,404.75) and (532.9,344.74) .. (592.99,330.49) ;
\draw [shift={(596.67,329.67)}, rotate = 528.3399999999999] [fill={rgb, 255:red, 208; green, 2; blue, 27 }  ,fill opacity=1 ][line width=0.08]  [draw opacity=0] (11.61,-5.58) -- (0,0) -- (11.61,5.58) -- cycle    ;
\draw [shift={(443.67,454.67)}, rotate = 294.08] [fill={rgb, 255:red, 208; green, 2; blue, 27 }  ,fill opacity=1 ][line width=0.08]  [draw opacity=0] (11.61,-5.58) -- (0,0) -- (11.61,5.58) -- cycle    ;
\draw [color={rgb, 255:red, 208; green, 2; blue, 27 }  ,draw opacity=1 ][line width=1.5]    (74.01,324.65) -- (227.1,473.86) ;
\draw [shift={(229.96,476.65)}, rotate = 224.27] [fill={rgb, 255:red, 208; green, 2; blue, 27 }  ,fill opacity=1 ][line width=0.08]  [draw opacity=0] (11.61,-5.58) -- (0,0) -- (11.61,5.58) -- cycle    ;
\draw [shift={(71.15,321.85)}, rotate = 44.27] [fill={rgb, 255:red, 208; green, 2; blue, 27 }  ,fill opacity=1 ][line width=0.08]  [draw opacity=0] (11.61,-5.58) -- (0,0) -- (11.61,5.58) -- cycle    ;
\draw [color={rgb, 255:red, 208; green, 2; blue, 27 }  ,draw opacity=1 ][line width=1.5]    (74.84,482.37) -- (228.3,324.96) ;
\draw [shift={(231.09,322.1)}, rotate = 494.27] [fill={rgb, 255:red, 208; green, 2; blue, 27 }  ,fill opacity=1 ][line width=0.08]  [draw opacity=0] (11.61,-5.58) -- (0,0) -- (11.61,5.58) -- cycle    ;
\draw [shift={(72.05,485.24)}, rotate = 314.27] [fill={rgb, 255:red, 208; green, 2; blue, 27 }  ,fill opacity=1 ][line width=0.08]  [draw opacity=0] (11.61,-5.58) -- (0,0) -- (11.61,5.58) -- cycle    ;
\draw [color={rgb, 255:red, 208; green, 2; blue, 27 }  ,draw opacity=1 ][line width=1.5]    (88.46,490.14) -- (210.17,490.14) ;
\draw [shift={(214.17,490.14)}, rotate = 180] [fill={rgb, 255:red, 208; green, 2; blue, 27 }  ,fill opacity=1 ][line width=0.08]  [draw opacity=0] (11.61,-5.58) -- (0,0) -- (11.61,5.58) -- cycle    ;
\draw [shift={(84.46,490.14)}, rotate = 0] [fill={rgb, 255:red, 208; green, 2; blue, 27 }  ,fill opacity=1 ][line width=0.08]  [draw opacity=0] (11.61,-5.58) -- (0,0) -- (11.61,5.58) -- cycle    ;
\draw [color={rgb, 255:red, 208; green, 2; blue, 27 }  ,draw opacity=1 ][line width=1.5]    (92.97,314.74) -- (214.68,314.74) ;
\draw [shift={(218.68,314.74)}, rotate = 180] [fill={rgb, 255:red, 208; green, 2; blue, 27 }  ,fill opacity=1 ][line width=0.08]  [draw opacity=0] (11.61,-5.58) -- (0,0) -- (11.61,5.58) -- cycle    ;
\draw [shift={(88.97,314.74)}, rotate = 0] [fill={rgb, 255:red, 208; green, 2; blue, 27 }  ,fill opacity=1 ][line width=0.08]  [draw opacity=0] (11.61,-5.58) -- (0,0) -- (11.61,5.58) -- cycle    ;
\draw [color={rgb, 255:red, 208; green, 2; blue, 27 }  ,draw opacity=1 ][line width=1.5]    (52.24,430.17) -- (52.64,375.87) ;
\draw [shift={(52.67,371.87)}, rotate = 450.42] [fill={rgb, 255:red, 208; green, 2; blue, 27 }  ,fill opacity=1 ][line width=0.08]  [draw opacity=0] (11.61,-5.58) -- (0,0) -- (11.61,5.58) -- cycle    ;
\draw [shift={(52.21,434.17)}, rotate = 270.42] [fill={rgb, 255:red, 208; green, 2; blue, 27 }  ,fill opacity=1 ][line width=0.08]  [draw opacity=0] (11.61,-5.58) -- (0,0) -- (11.61,5.58) -- cycle    ;
\draw [color={rgb, 255:red, 65; green, 117; blue, 5 }  ,draw opacity=1 ][line width=1.5]    (574.87,365.97) -- (618.67,324.84) ;
\draw [shift={(621.58,322.1)}, rotate = 496.8] [fill={rgb, 255:red, 65; green, 117; blue, 5 }  ,fill opacity=1 ][line width=0.08]  [draw opacity=0] (11.61,-5.58) -- (0,0) -- (11.61,5.58) -- cycle    ;
\draw [shift={(571.95,368.71)}, rotate = 316.8] [fill={rgb, 255:red, 65; green, 117; blue, 5 }  ,fill opacity=1 ][line width=0.08]  [draw opacity=0] (11.61,-5.58) -- (0,0) -- (11.61,5.58) -- cycle    ;
\draw [color={rgb, 255:red, 65; green, 117; blue, 5 }  ,draw opacity=1 ][line width=1.5]    (441.45,478.51) -- (479.12,434.31) ;
\draw [shift={(481.71,431.27)}, rotate = 490.44] [fill={rgb, 255:red, 65; green, 117; blue, 5 }  ,fill opacity=1 ][line width=0.08]  [draw opacity=0] (11.61,-5.58) -- (0,0) -- (11.61,5.58) -- cycle    ;
\draw [shift={(438.85,481.56)}, rotate = 310.44] [fill={rgb, 255:red, 65; green, 117; blue, 5 }  ,fill opacity=1 ][line width=0.08]  [draw opacity=0] (11.61,-5.58) -- (0,0) -- (11.61,5.58) -- cycle    ;
\draw [color={rgb, 255:red, 65; green, 117; blue, 5 }  ,draw opacity=1 ][line width=1.5]    (480.02,365.88) -- (441.68,327.39) ;
\draw [shift={(438.85,324.55)}, rotate = 405.11] [fill={rgb, 255:red, 65; green, 117; blue, 5 }  ,fill opacity=1 ][line width=0.08]  [draw opacity=0] (11.61,-5.58) -- (0,0) -- (11.61,5.58) -- cycle    ;
\draw [shift={(482.84,368.71)}, rotate = 225.11] [fill={rgb, 255:red, 65; green, 117; blue, 5 }  ,fill opacity=1 ][line width=0.08]  [draw opacity=0] (11.61,-5.58) -- (0,0) -- (11.61,5.58) -- cycle    ;
\draw [color={rgb, 255:red, 65; green, 117; blue, 5 }  ,draw opacity=1 ][line width=1.5]    (616.55,461.22) -- (573.6,416.69) ;
\draw [shift={(570.82,413.81)}, rotate = 406.03999999999996] [fill={rgb, 255:red, 65; green, 117; blue, 5 }  ,fill opacity=1 ][line width=0.08]  [draw opacity=0] (11.61,-5.58) -- (0,0) -- (11.61,5.58) -- cycle    ;
\draw [shift={(619.33,464.1)}, rotate = 226.04] [fill={rgb, 255:red, 65; green, 117; blue, 5 }  ,fill opacity=1 ][line width=0.08]  [draw opacity=0] (11.61,-5.58) -- (0,0) -- (11.61,5.58) -- cycle    ;
\draw [color={rgb, 255:red, 208; green, 2; blue, 27 }  ,draw opacity=1 ][line width=1.5]    (472.97,314.74) -- (594.68,314.74) ;
\draw [shift={(598.68,314.74)}, rotate = 180] [fill={rgb, 255:red, 208; green, 2; blue, 27 }  ,fill opacity=1 ][line width=0.08]  [draw opacity=0] (11.61,-5.58) -- (0,0) -- (11.61,5.58) -- cycle    ;
\draw [shift={(468.97,314.74)}, rotate = 0] [fill={rgb, 255:red, 208; green, 2; blue, 27 }  ,fill opacity=1 ][line width=0.08]  [draw opacity=0] (11.61,-5.58) -- (0,0) -- (11.61,5.58) -- cycle    ;
\draw [color={rgb, 255:red, 208; green, 2; blue, 27 }  ,draw opacity=1 ][line width=1.5]    (469.97,486.74) -- (591.68,486.74) ;
\draw [shift={(595.68,486.74)}, rotate = 180] [fill={rgb, 255:red, 208; green, 2; blue, 27 }  ,fill opacity=1 ][line width=0.08]  [draw opacity=0] (11.61,-5.58) -- (0,0) -- (11.61,5.58) -- cycle    ;
\draw [shift={(465.97,486.74)}, rotate = 0] [fill={rgb, 255:red, 208; green, 2; blue, 27 }  ,fill opacity=1 ][line width=0.08]  [draw opacity=0] (11.61,-5.58) -- (0,0) -- (11.61,5.58) -- cycle    ;
\draw  [fill={rgb, 255:red, 185; green, 185; blue, 185 }  ,fill opacity=1 ][line width=1.5]  (478.56,378.57) .. controls (478.56,370.96) and (484.73,364.79) .. (492.34,364.79) -- (561.55,364.79) .. controls (569.16,364.79) and (575.33,370.96) .. (575.33,378.57) -- (575.33,419.93) .. controls (575.33,427.55) and (569.16,433.72) .. (561.55,433.72) -- (492.34,433.72) .. controls (484.73,433.72) and (478.56,427.55) .. (478.56,419.93) -- cycle ;
\draw [color={rgb, 255:red, 208; green, 2; blue, 27 }  ,draw opacity=1 ][line width=1.5]    (247.24,428.17) -- (247.64,373.87) ;
\draw [shift={(247.67,369.87)}, rotate = 450.42] [fill={rgb, 255:red, 208; green, 2; blue, 27 }  ,fill opacity=1 ][line width=0.08]  [draw opacity=0] (11.61,-5.58) -- (0,0) -- (11.61,5.58) -- cycle    ;
\draw [shift={(247.21,432.17)}, rotate = 270.42] [fill={rgb, 255:red, 208; green, 2; blue, 27 }  ,fill opacity=1 ][line width=0.08]  [draw opacity=0] (11.61,-5.58) -- (0,0) -- (11.61,5.58) -- cycle    ;
\draw [color={rgb, 255:red, 208; green, 2; blue, 27 }  ,draw opacity=1 ][line width=1.5]    (428.24,433.17) -- (428.64,378.87) ;
\draw [shift={(428.67,374.87)}, rotate = 450.42] [fill={rgb, 255:red, 208; green, 2; blue, 27 }  ,fill opacity=1 ][line width=0.08]  [draw opacity=0] (11.61,-5.58) -- (0,0) -- (11.61,5.58) -- cycle    ;
\draw [shift={(428.21,437.17)}, rotate = 270.42] [fill={rgb, 255:red, 208; green, 2; blue, 27 }  ,fill opacity=1 ][line width=0.08]  [draw opacity=0] (11.61,-5.58) -- (0,0) -- (11.61,5.58) -- cycle    ;
\draw [color={rgb, 255:red, 208; green, 2; blue, 27 }  ,draw opacity=1 ][line width=1.5]    (629.51,430.32) -- (629.9,376.02) ;
\draw [shift={(629.93,372.02)}, rotate = 450.42] [fill={rgb, 255:red, 208; green, 2; blue, 27 }  ,fill opacity=1 ][line width=0.08]  [draw opacity=0] (11.61,-5.58) -- (0,0) -- (11.61,5.58) -- cycle    ;
\draw [shift={(629.48,434.32)}, rotate = 270.42] [fill={rgb, 255:red, 208; green, 2; blue, 27 }  ,fill opacity=1 ][line width=0.08]  [draw opacity=0] (11.61,-5.58) -- (0,0) -- (11.61,5.58) -- cycle    ;
\draw  [fill={rgb, 255:red, 185; green, 185; blue, 185 }  ,fill opacity=1 ][line width=1.5]  (298.56,130.21) .. controls (298.56,122.6) and (304.73,116.43) .. (312.34,116.43) -- (381.55,116.43) .. controls (389.16,116.43) and (395.33,122.6) .. (395.33,130.21) -- (395.33,171.58) .. controls (395.33,179.19) and (389.16,185.36) .. (381.55,185.36) -- (312.34,185.36) .. controls (304.73,185.36) and (298.56,179.19) .. (298.56,171.58) -- cycle ;
\draw (117.27,147.84) node  {\includegraphics[width=84.6pt,height=31.86pt]{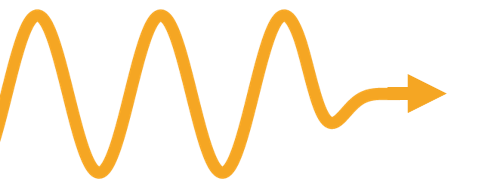}};
\draw (509.38,149.13) node  {\includegraphics[width=98.85pt,height=31.26pt]{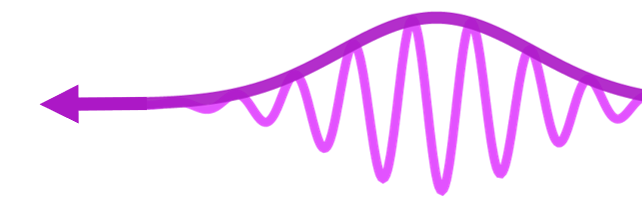}};
\draw [line width=1.5]    (248.4,115.09) -- (248.4,52.09) ;
\draw [shift={(248.4,48.09)}, rotate = 450] [fill={rgb, 255:red, 0; green, 0; blue, 0 }  ][line width=0.08]  [draw opacity=0] (13.4,-6.43) -- (0,0) -- (13.4,6.44) -- (8.9,0) -- cycle    ;
\draw  [fill={rgb, 255:red, 74; green, 168; blue, 226 }  ,fill opacity=1 ] (238.56,84.64) .. controls (238.56,78.34) and (243.26,73.23) .. (249.05,73.23) .. controls (254.84,73.23) and (259.54,78.34) .. (259.54,84.64) .. controls (259.54,90.94) and (254.84,96.05) .. (249.05,96.05) .. controls (243.26,96.05) and (238.56,90.94) .. (238.56,84.64) -- cycle ;

\draw [line width=1.5]    (249.4,230.09) -- (249.4,167.09) ;
\draw [shift={(249.4,163.09)}, rotate = 450] [fill={rgb, 255:red, 0; green, 0; blue, 0 }  ][line width=0.08]  [draw opacity=0] (13.4,-6.43) -- (0,0) -- (13.4,6.44) -- (8.9,0) -- cycle    ;
\draw  [fill={rgb, 255:red, 74; green, 168; blue, 226 }  ,fill opacity=1 ] (239.56,199.64) .. controls (239.56,193.34) and (244.26,188.23) .. (250.05,188.23) .. controls (255.84,188.23) and (260.54,193.34) .. (260.54,199.64) .. controls (260.54,205.94) and (255.84,211.05) .. (250.05,211.05) .. controls (244.26,211.05) and (239.56,205.94) .. (239.56,199.64) -- cycle ;

\draw [line width=1.5]    (436.4,115.09) -- (436.4,52.09) ;
\draw [shift={(436.4,48.09)}, rotate = 450] [fill={rgb, 255:red, 0; green, 0; blue, 0 }  ][line width=0.08]  [draw opacity=0] (13.4,-6.43) -- (0,0) -- (13.4,6.44) -- (8.9,0) -- cycle    ;
\draw  [fill={rgb, 255:red, 74; green, 168; blue, 226 }  ,fill opacity=1 ] (426.56,84.64) .. controls (426.56,78.34) and (431.26,73.23) .. (437.05,73.23) .. controls (442.84,73.23) and (447.54,78.34) .. (447.54,84.64) .. controls (447.54,90.94) and (442.84,96.05) .. (437.05,96.05) .. controls (431.26,96.05) and (426.56,90.94) .. (426.56,84.64) -- cycle ;

\draw [line width=1.5]    (437.4,231.09) -- (437.4,168.09) ;
\draw [shift={(437.4,164.09)}, rotate = 450] [fill={rgb, 255:red, 0; green, 0; blue, 0 }  ][line width=0.08]  [draw opacity=0] (13.4,-6.43) -- (0,0) -- (13.4,6.44) -- (8.9,0) -- cycle    ;
\draw  [fill={rgb, 255:red, 74; green, 168; blue, 226 }  ,fill opacity=1 ] (427.56,200.64) .. controls (427.56,194.34) and (432.26,189.23) .. (438.05,189.23) .. controls (443.84,189.23) and (448.54,194.34) .. (448.54,200.64) .. controls (448.54,206.94) and (443.84,212.05) .. (438.05,212.05) .. controls (432.26,212.05) and (427.56,206.94) .. (427.56,200.64) -- cycle ;

\draw [line width=1.5]    (51.4,347.21) -- (51.4,284.21) ;
\draw [shift={(51.4,280.21)}, rotate = 450] [fill={rgb, 255:red, 0; green, 0; blue, 0 }  ][line width=0.08]  [draw opacity=0] (13.4,-6.43) -- (0,0) -- (13.4,6.44) -- (8.9,0) -- cycle    ;
\draw  [fill={rgb, 255:red, 74; green, 168; blue, 226 }  ,fill opacity=1 ] (41.56,316.76) .. controls (41.56,310.46) and (46.26,305.35) .. (52.05,305.35) .. controls (57.84,305.35) and (62.54,310.46) .. (62.54,316.76) .. controls (62.54,323.06) and (57.84,328.17) .. (52.05,328.17) .. controls (46.26,328.17) and (41.56,323.06) .. (41.56,316.76) -- cycle ;

\draw [line width=1.5]    (52.4,520.21) -- (52.4,457.21) ;
\draw [shift={(52.4,453.21)}, rotate = 450] [fill={rgb, 255:red, 0; green, 0; blue, 0 }  ][line width=0.08]  [draw opacity=0] (13.4,-6.43) -- (0,0) -- (13.4,6.44) -- (8.9,0) -- cycle    ;
\draw  [fill={rgb, 255:red, 74; green, 168; blue, 226 }  ,fill opacity=1 ] (42.56,489.76) .. controls (42.56,483.46) and (47.26,478.35) .. (53.05,478.35) .. controls (58.84,478.35) and (63.54,483.46) .. (63.54,489.76) .. controls (63.54,496.06) and (58.84,501.17) .. (53.05,501.17) .. controls (47.26,501.17) and (42.56,496.06) .. (42.56,489.76) -- cycle ;

\draw [line width=1.5]    (248.4,515.21) -- (248.4,452.21) ;
\draw [shift={(248.4,448.21)}, rotate = 450] [fill={rgb, 255:red, 0; green, 0; blue, 0 }  ][line width=0.08]  [draw opacity=0] (13.4,-6.43) -- (0,0) -- (13.4,6.44) -- (8.9,0) -- cycle    ;
\draw  [fill={rgb, 255:red, 74; green, 168; blue, 226 }  ,fill opacity=1 ] (238.56,484.76) .. controls (238.56,478.46) and (243.26,473.35) .. (249.05,473.35) .. controls (254.84,473.35) and (259.54,478.46) .. (259.54,484.76) .. controls (259.54,491.06) and (254.84,496.17) .. (249.05,496.17) .. controls (243.26,496.17) and (238.56,491.06) .. (238.56,484.76) -- cycle ;

\draw [line width=1.5]    (247.4,348.21) -- (247.4,285.21) ;
\draw [shift={(247.4,281.21)}, rotate = 450] [fill={rgb, 255:red, 0; green, 0; blue, 0 }  ][line width=0.08]  [draw opacity=0] (13.4,-6.43) -- (0,0) -- (13.4,6.44) -- (8.9,0) -- cycle    ;
\draw  [fill={rgb, 255:red, 74; green, 168; blue, 226 }  ,fill opacity=1 ] (237.56,317.76) .. controls (237.56,311.46) and (242.26,306.35) .. (248.05,306.35) .. controls (253.84,306.35) and (258.54,311.46) .. (258.54,317.76) .. controls (258.54,324.06) and (253.84,329.17) .. (248.05,329.17) .. controls (242.26,329.17) and (237.56,324.06) .. (237.56,317.76) -- cycle ;

\draw [line width=1.5]    (428.4,345.21) -- (428.4,282.21) ;
\draw [shift={(428.4,278.21)}, rotate = 450] [fill={rgb, 255:red, 0; green, 0; blue, 0 }  ][line width=0.08]  [draw opacity=0] (13.4,-6.43) -- (0,0) -- (13.4,6.44) -- (8.9,0) -- cycle    ;
\draw  [fill={rgb, 255:red, 74; green, 168; blue, 226 }  ,fill opacity=1 ] (418.56,314.76) .. controls (418.56,308.46) and (423.26,303.35) .. (429.05,303.35) .. controls (434.84,303.35) and (439.54,308.46) .. (439.54,314.76) .. controls (439.54,321.06) and (434.84,326.17) .. (429.05,326.17) .. controls (423.26,326.17) and (418.56,321.06) .. (418.56,314.76) -- cycle ;

\draw [line width=1.5]    (428.4,517.21) -- (428.4,454.21) ;
\draw [shift={(428.4,450.21)}, rotate = 450] [fill={rgb, 255:red, 0; green, 0; blue, 0 }  ][line width=0.08]  [draw opacity=0] (13.4,-6.43) -- (0,0) -- (13.4,6.44) -- (8.9,0) -- cycle    ;
\draw  [fill={rgb, 255:red, 74; green, 168; blue, 226 }  ,fill opacity=1 ] (418.56,486.76) .. controls (418.56,480.46) and (423.26,475.35) .. (429.05,475.35) .. controls (434.84,475.35) and (439.54,480.46) .. (439.54,486.76) .. controls (439.54,493.06) and (434.84,498.17) .. (429.05,498.17) .. controls (423.26,498.17) and (418.56,493.06) .. (418.56,486.76) -- cycle ;

\draw [line width=1.5]    (628.4,346.21) -- (628.4,283.21) ;
\draw [shift={(628.4,279.21)}, rotate = 450] [fill={rgb, 255:red, 0; green, 0; blue, 0 }  ][line width=0.08]  [draw opacity=0] (13.4,-6.43) -- (0,0) -- (13.4,6.44) -- (8.9,0) -- cycle    ;
\draw  [fill={rgb, 255:red, 74; green, 168; blue, 226 }  ,fill opacity=1 ] (618.56,315.76) .. controls (618.56,309.46) and (623.26,304.35) .. (629.05,304.35) .. controls (634.84,304.35) and (639.54,309.46) .. (639.54,315.76) .. controls (639.54,322.06) and (634.84,327.17) .. (629.05,327.17) .. controls (623.26,327.17) and (618.56,322.06) .. (618.56,315.76) -- cycle ;

\draw [line width=1.5]    (630.4,516.21) -- (630.4,453.21) ;
\draw [shift={(630.4,449.21)}, rotate = 450] [fill={rgb, 255:red, 0; green, 0; blue, 0 }  ][line width=0.08]  [draw opacity=0] (13.4,-6.43) -- (0,0) -- (13.4,6.44) -- (8.9,0) -- cycle    ;
\draw  [fill={rgb, 255:red, 74; green, 168; blue, 226 }  ,fill opacity=1 ] (620.56,485.76) .. controls (620.56,479.46) and (625.26,474.35) .. (631.05,474.35) .. controls (636.84,474.35) and (641.54,479.46) .. (641.54,485.76) .. controls (641.54,492.06) and (636.84,497.17) .. (631.05,497.17) .. controls (625.26,497.17) and (620.56,492.06) .. (620.56,485.76) -- cycle ;

\draw (388,269.36) node [anchor=north west][inner sep=0.75pt]   [align=left] {c)};
\draw (480.12,386.24) node [anchor=north west][inner sep=0.75pt]  [color={rgb, 255:red, 0; green, 0; blue, 0 }  ,opacity=1 ] [align=left] {Coupler };
\draw (77.81,509.68) node [anchor=north west][inner sep=0.75pt]  [color={rgb, 255:red, 0; green, 0; blue, 0 }  ,opacity=1 ] [align=left] {Coupler OFF };
\draw (461.19,501.22) node [anchor=north west][inner sep=0.75pt]  [color={rgb, 255:red, 0; green, 0; blue, 0 }  ,opacity=1 ] [align=left] {Coupler ON };
\draw (8.17,264.13) node [anchor=north west][inner sep=0.75pt]   [align=left] {b)};
\draw (111.87,272.27) node [anchor=north west][inner sep=0.75pt]    {$J_{Z}^{( ij)} ,J_{X}^{( ij)}$};
\draw (516.87,322.27) node [anchor=north west][inner sep=0.75pt]    {$M$};
\draw (7,16) node [anchor=north west][inner sep=0.75pt]   [align=left] {a)};
\draw (300.12,137.89) node [anchor=north west][inner sep=0.75pt]  [color={rgb, 255:red, 0; green, 0; blue, 0 }  ,opacity=1 ] [align=left] {Coupler };
\draw (279.46,36.55) node [anchor=north west][inner sep=0.75pt]  [color={rgb, 255:red, 0; green, 0; blue, 0 }  ,opacity=1 ] [align=left] {Spin System};
\draw (40.87,52.46) node [anchor=north west][inner sep=0.75pt]   [align=left] {\begin{minipage}[lt]{45.8pt}\setlength\topsep{0pt}
\begin{center}
Static\\Detection
\end{center}

\end{minipage}};
\draw (492.87,53.79) node [anchor=north west][inner sep=0.75pt]   [align=left] {\begin{minipage}[lt]{45.8pt}\setlength\topsep{0pt}
\begin{center}
Dynamic \\Detection
\end{center}

\end{minipage}};

\end{tikzpicture}

    \caption{System of artificial spins coupled via a multi-spin coupler. (a) The types of techniques to detect multi-spin interactions considered in this work. Static detection techniques probe steady state behavior with continuous-wave drives, whereas dynamic techniques such as coupler pulses lead to transient behavior. (b) Schematic of the system for $n=4$ when the coupler is off, including pairwise interactions $J_Z,J_X$ between all pairs of spins. (c) When the coupler is turned on, a $4$-local term with coupling strength $M$ as well as spurious terms of lower locality are added to the system Hamiltonian.}
    \label{fig:coupler}
\end{figure}
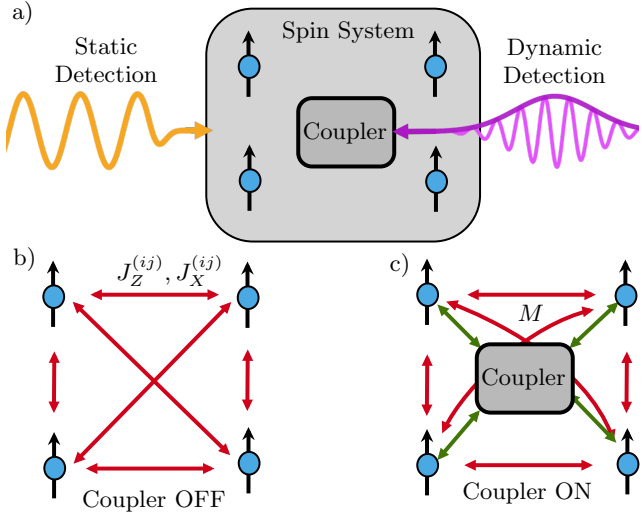

\textit{Coupled multi-spin Hamiltonians}.---We consider an $n$-spin system in which all subsets of spins are coupled. 
Additionally, we consider that the system has a \textit{coupler} as portrayed in \figref{fig:coupler}.
We assume that the coupler can be turned on and off.
When it is on, it adds an $n$-local term to the Hamiltonian that couples all spins and takes the form $M Z_1\otimes Z_2 \otimes \ldots \otimes Z_n$ in the system Hamiltonian.
The $n$-local coupling strength is given by $M$, and $Z_i$ is the Pauli $Z$ matrix for the $i$-th spin.

The following $n$-spin Hamiltonian family is considered:
\begin{equation}
\label{eq:hamiltonian}
    H = \sum_{s\in S} \delta_s X_s + \epsilon_s Z_s + \sum_{Q\subseteq S, |Q|>1} J_{Z}^{(Q)} \Motimes_{s\in Q} Z_{s} + J_{X}^{(Q)} \Motimes_{s\in Q} X_{s},
\end{equation}
where $X_i,Z_i$ are the $X$ and $Z$ Pauli matrices, respectively.
The first sum is over individual spins in the set $S$ of all $n$ spins, and the second runs over all non-empty subsets $Q$ of size larger than one. 
For each spin subset $Q$, the Hamiltonian defines a coupling between all spins in the subset, which can be a $Z$-type coupling with strength $J_Z^{(Q)}$ or an $X$-type coupling with strength $J_X^{(Q)}$.
We only consider $Z$-coupling and transverse $X$ terms, but our techniques can be generalized to include $Y$-type coupling.
Inspired by the relevant platform of single-loop flux qubits \cite{Orlando_1999,You_2007,yan2016flux}, we assume that the single spin $Z$-fields $\epsilon_s$ are the only tunable parameters in this Hamiltonian apart from the ability to turn the coupler on and off.
We can vary each $\epsilon_s$ parameter for each spin $s$ individually in the range from $0$ to $\epsilon_{\max}$. 
Similarly, we label by $\delta_s$ the single spin $X$-field, which is not tunable.
The $n$-local term of interest is present when $Q = S$. 
We assume the $n$-local term along the $X$ direction to be fixed to $J_X^{(S)} = 0$ and we define $J_Z^{(S)} = M$.
For the system model, we use parameters that are experimentally feasible for coupled superconducting flux qubits, a promising platform for realizing artificial spin systems.
We set $\delta_s = 2\pi\times 2\ \text{GHz}, \epsilon_{\max} = 2\pi\times 10\ \text{GHz}$ for the single-qubit fields, and for the coupling parameters of locality lower than $n$ we assume $J_X^{(Q)},J_Z^{(Q)}$ sampled in the range of $2\pi\times [0,300)\ \text{MHz}$ \cite{Weber_2017,Harris_2007}.
Finally, for the $n$-local term, we base our values on the practical coupler proposal discussed in Ref.~\cite{menke2019automated}, which estimates a 4-local coupling of about $2\pi\times 500\ \text{MHz}$. 
However, we expect that the first experimental prototypes will exhibit a lower coupling and assume $M = 2\pi\times 50 \ \text{MHz}$.
Although these Hamiltonian parameters are aimed to highlight experimental feasibility, we emphasize that our techniques are also applicable to other systems encoding artificial spins.

Throughout this work, we denote a $k$-local coupling term as an additive term in the Hamiltonian of the form $Z_1\otimes Z_2\otimes\ldots\otimes Z_k$ (or along the $X$ direction) between any group of $k$ spins. Similarly, we define a $k$-local Hamiltonian as including $l$-local coupling terms from $l=1$ to $l=k$. The \textit{locality} of such a $k$-local Hamiltonian is the quantity $k$.

Most practical $n$-local coupler setups, as in Refs. \cite{Schondorf_2019,melanson2019tunable,menke2019automated}, will be imperfect:
Whenever the coupler is turned on, it generates spurious terms, inadvertently altering all other Hamiltonian parameters -- except the $n$-local parameters $J_X^{(S)}$ and $J_Z^{(S)}$ --  by a value in the range $\pm \eta M$, where $\eta$ is the relative spurious term amplitude. Here we consider $\eta$ fixed at $\eta = 1/2$, but generalize to include larger values of $\eta$ in the Supplementary Information.

\begin{figure}[t]
\centering
\begin{tikzpicture}[x=0.75pt,y=0.75pt,yscale=-1,xscale=1]

\draw (136,107.84) node  {\includegraphics[width=184.5pt,height=141.72pt]{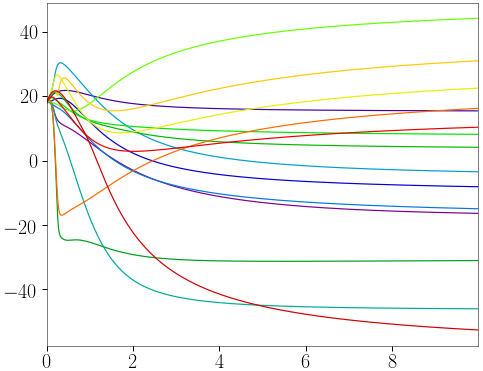}};
\draw  [draw opacity=0][fill={rgb, 255:red, 255; green, 255; blue, 255 }  ,fill opacity=1 ] (264,17.36) -- (344,17.36) -- (344,30.36) -- (264,30.36) -- cycle ;
\draw (294,111.1) node  {\includegraphics[width=31.5pt,height=119.62pt]{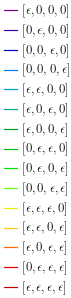}};

\draw (83.55,203.2) node [anchor=north west][inner sep=0.75pt]  [font=\footnotesize] [align=left] {$\displaystyle Z$-field Strength $\displaystyle \varepsilon $ (GHz) };
\draw (4.03,197.72) node [anchor=north west][inner sep=0.75pt]  [font=\footnotesize,rotate=-270.95] [align=left] {Transition energy variation (MHz)};
\draw (259.8,10) node [anchor=north west][inner sep=0.75pt]  [font=\scriptsize] [align=center] {\begin{minipage}[lt]{45.16pt}\setlength\topsep{0pt}
\begin{center}
Field \\Configuration
\end{center}

\end{minipage}};

\end{tikzpicture}

    \caption{Transition energy variation of a 4-spin system. It is shown how the transition energy variation changes with the local $Z$-field $\epsilon$ for all field configurations. The legend indicates to which of the $4$ spins the field is applied for each curve.}
    \label{fig:spaghetti}
\end{figure}
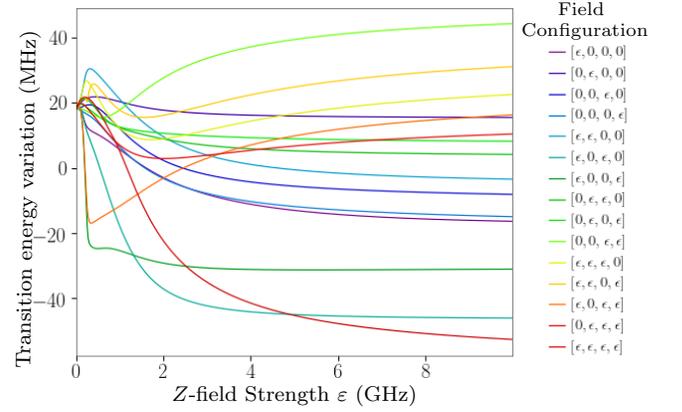

\textit{Static detection technique}.---The first proposed detection technique is based on analyzing variations in the transition energy of the system both with the coupler on and off.
The detection principle is based on the fact that these variations contain an $n$-local signature, which can only be generated by a Hamiltonian with the desired $n$-local term, and not with an $(n-1)$-local Hamiltonian.
The proposed experimental procedure consists of measuring the transition energy first with the coupler off $E_{01}^{(0)}$, then on $E_{01}^{(M)}$, and analyzing the variation of the transition energy:

\begin{equation}
    \{\Delta E^{M}\} =\{ E_{01}^{(M)}-E_{01}^{(0)}\}
\end{equation}

\noindent

The transition energy variation $\Delta E^{M}$ is determined as a function of the tunable $Z$-fields $\epsilon_i$ while maintaining some of the spins fixed at $\epsilon_i = 0$.
Specifically, for each subset of spins $Q\subseteq S$, we set $\epsilon_i = \epsilon \ \forall i\in Q$ and measure the transition energy variation as a function of $\epsilon$, varying from $0$ to $\epsilon_{\max}$.
Doing so for all subsets of spins, we calculate the transition energy variation as a function of $\epsilon$ for $2^n-1$ distinct configurations, as visualized in \figref{fig:spaghetti}. Many of these curves do not simply collapse onto each other due to the sampled spurious terms, which break the system degeneracy.

A central claim that we make is that the spectroscopy curves contain an $n$-local \textit{signature}, which can only be explained by an $n$-local Hamiltonian and by none other with lower locality.
We find the signature via a pair of Hamiltonian model fits based on \eqref{eq:hamiltonian}. Specifically, we assume that we know all original coupling parameters with the coupler off.
Experimental techniques exist to characterize such 1- and 2-local terms reliably \cite{Harris_2007,Weber_2017,Ozfidan_2020,krantz2019quantum}.
Using a least squares fitting procedure \cite{2020SciPy-NMeth}, one fit is then performed with an $n$-local Hamiltonian and another with an $(n-1)$-local Hamiltonian.
The fits determine all spurious shifts and the $n$-local term.
By comparing the two fits via a fit quality measure, we can differentiate whether the system is described by an $n$-local Hamiltonian or one of lower locality. 
Fewer than the $2^n-1$ spin configurations can be used for the curve fits or the $\epsilon$ parameters can be chosen differently for different spins.
We do not analyze such modifications here, but they could potentially result in more scalable and efficient detection methodologies.

To analyze the performance of the fits on the $n$-local Hamiltonian data, we introduce a quality measure.
The first measure is the mean deviation between the fitted curve and the data points corresponding to the $n$-local Hamiltonian data.
Under the assumption of no experimental error, this quality measure easily distinguishes between an $n$-local and an $(n-1)$-local Hamiltonian model, concluding that the system Hamiltonian indeed includes an $n$-local term.
To study the procedural robustness, we analyze whether we can still distinguish the two Hamiltonian models in the presence of noise.

\begin{figure*}[ht]
\begin{tikzpicture}[x=0.75pt,y=0.75pt,yscale=-1,xscale=1]

\draw (488.37,227.79) node  {\includegraphics[width=238.35pt,height=296.09pt]{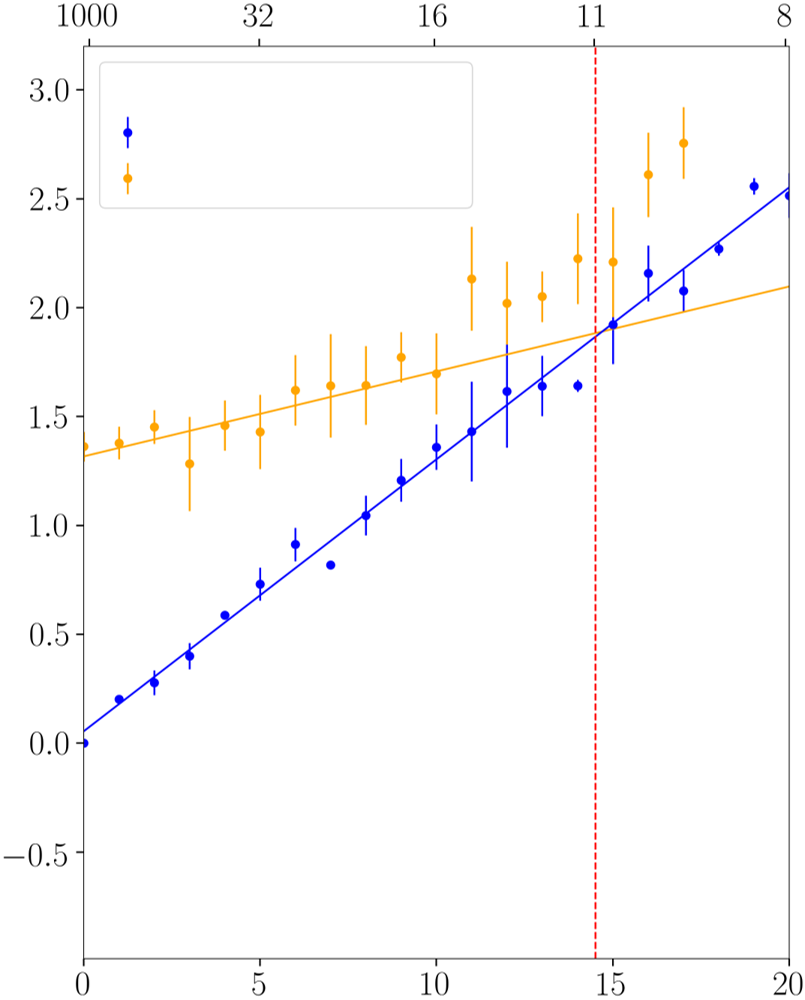}};

\draw  [fill={rgb, 255:red, 155; green, 155; blue, 155 }  ,fill opacity=1 ] (316.67,29.21) -- (400.39,29.21) -- (400.39,38.27) -- (395.86,38.27) -- (402.56,47.21) -- (409.27,38.27) -- (404.74,38.27) -- (404.74,24.87) -- (316.67,24.87) -- cycle ;
\draw  [fill={rgb, 255:red, 155; green, 155; blue, 155 }  ,fill opacity=1 ] (317.27,439.41) -- (615.03,439.41) -- (615.03,423.9) -- (607.28,423.9) -- (617.54,410.21) -- (627.8,423.9) -- (620.04,423.9) -- (620.04,444.42) -- (317.27,444.42) -- cycle ;
\draw (549.27,331.98) node  {\includegraphics[width=133.5pt,height=110.95pt]{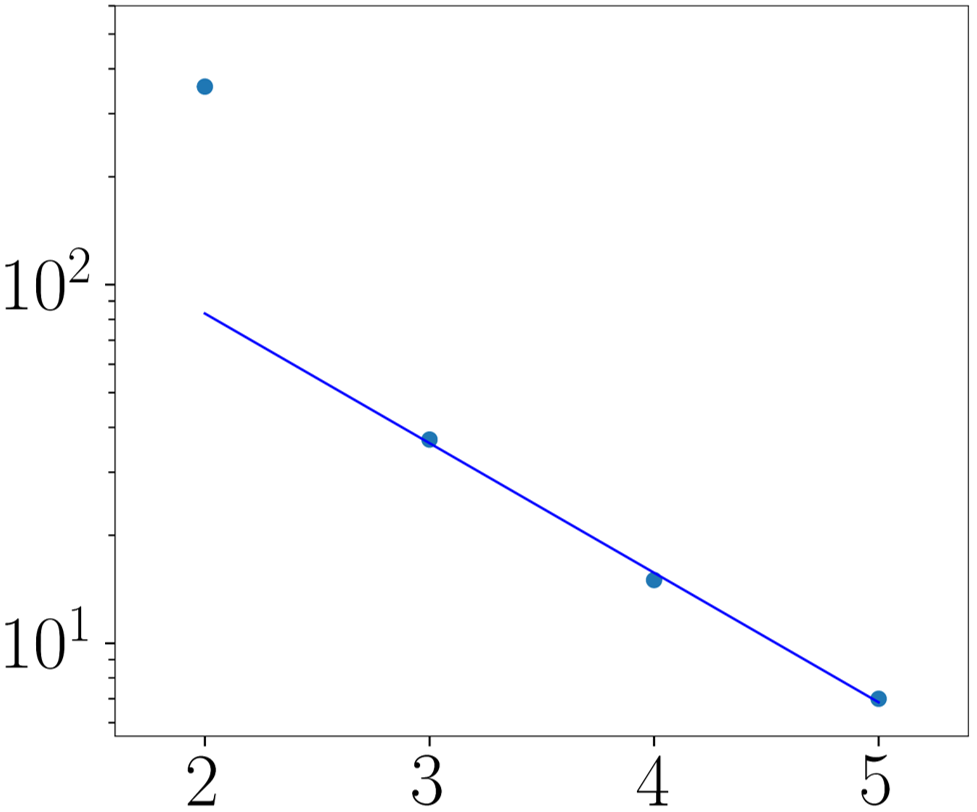}};
\draw  [color={rgb, 255:red, 255; green, 255; blue, 255 }  ,draw opacity=1 ][fill={rgb, 255:red, 255; green, 255; blue, 255 }  ,fill opacity=1 ] (206,374.44) -- (294.19,374.44) -- (294.19,420.95) -- (206,420.95) -- cycle ;
\draw  [color={rgb, 255:red, 255; green, 255; blue, 255 }  ,draw opacity=1 ][fill={rgb, 255:red, 255; green, 255; blue, 255 }  ,fill opacity=1 ] (186,359.44) -- (283.19,359.44) -- (283.19,375.95) -- (186,375.95) -- cycle ;
\draw  [color={rgb, 255:red, 255; green, 255; blue, 255 }  ,draw opacity=1 ][fill={rgb, 255:red, 255; green, 255; blue, 255 }  ,fill opacity=1 ] (180.1,357.2) -- (293.19,357.2) -- (293.19,416.04) -- (180.1,416.04) -- cycle ;
\draw  [color={rgb, 255:red, 255; green, 255; blue, 255 }  ,draw opacity=1 ][fill={rgb, 255:red, 255; green, 255; blue, 255 }  ,fill opacity=1 ] (171,341.94) -- (278.19,341.94) -- (278.19,357.54) -- (171,357.54) -- cycle ;
\draw (165.63,337.87) node  {\includegraphics[width=212.44pt,height=155.31pt]{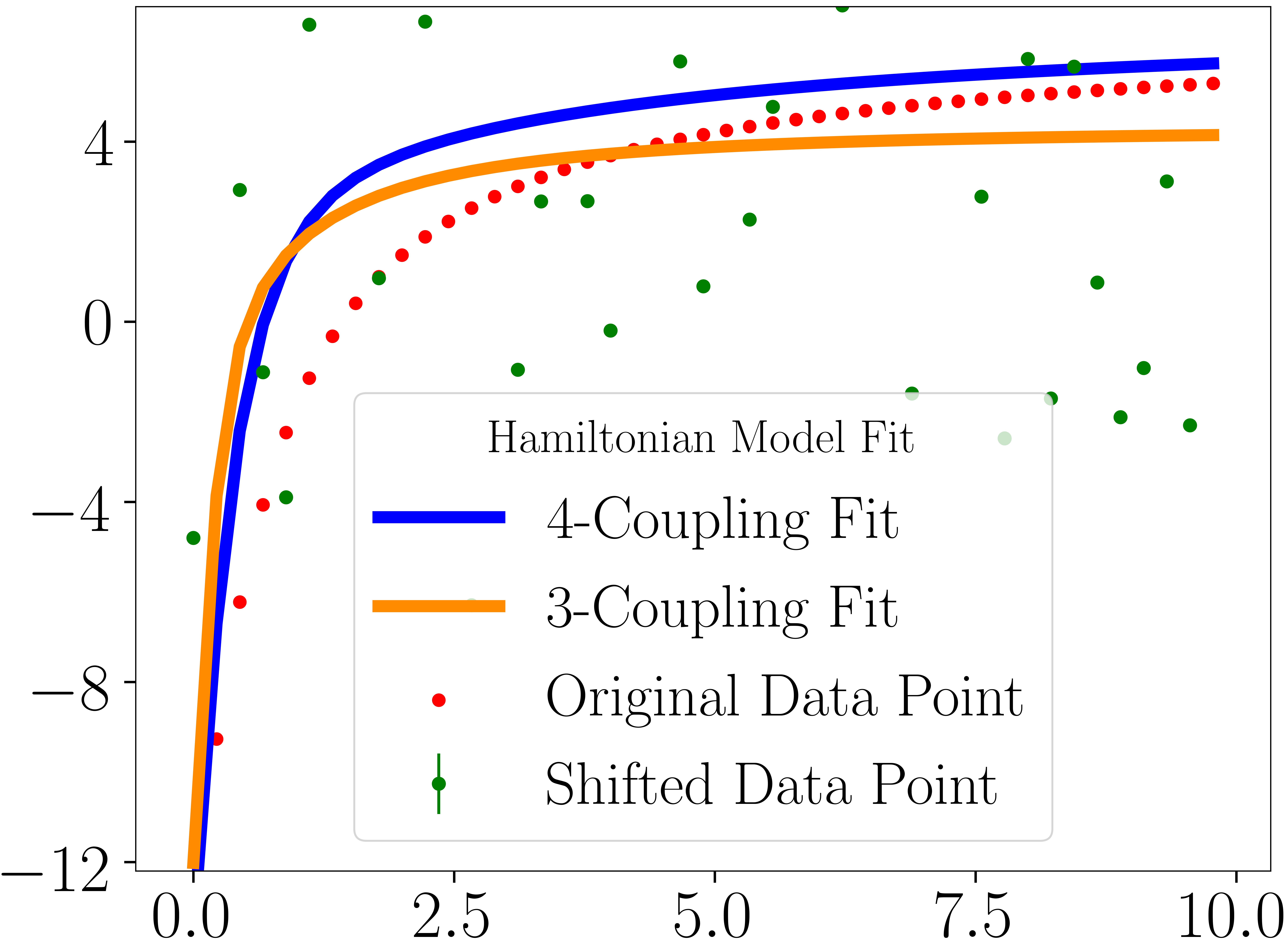}};
\draw  [color={rgb, 255:red, 255; green, 255; blue, 255 }  ,draw opacity=1 ][fill={rgb, 255:red, 255; green, 255; blue, 255 }  ,fill opacity=1 ] (142.67,337.58) -- (251.26,337.58) -- (251.26,414.92) -- (142.67,414.92) -- cycle ;
\draw  [color={rgb, 255:red, 255; green, 255; blue, 255 }  ,draw opacity=1 ][fill={rgb, 255:red, 255; green, 255; blue, 255 }  ,fill opacity=1 ] (129.33,324.24) -- (231.92,324.24) -- (231.92,340.92) -- (129.33,340.92) -- cycle ;

\draw (163.96,132.44) node  {\includegraphics[width=209.94pt,height=152.84pt]{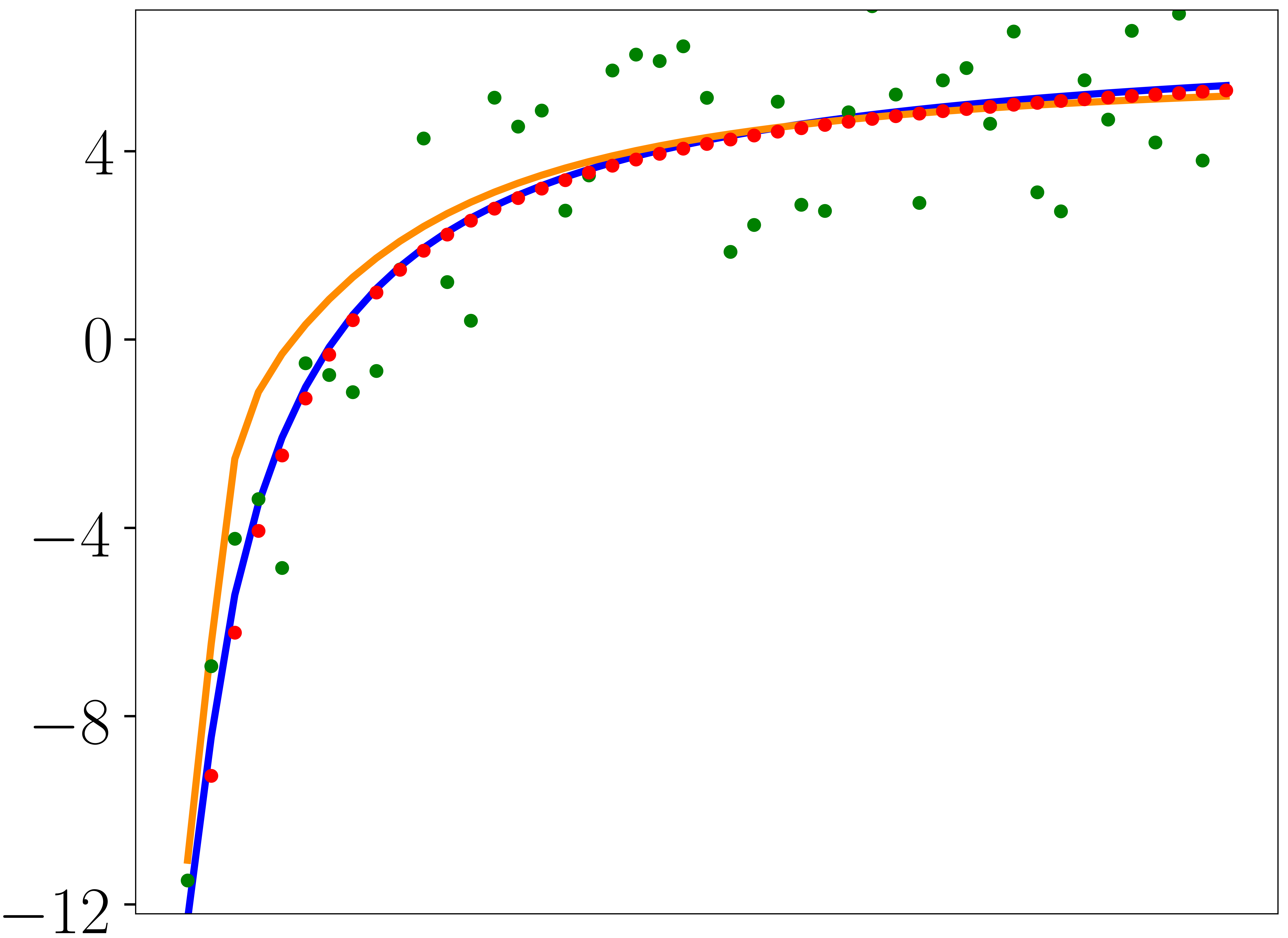}};

\draw (20.47,18.4) node [anchor=north west][inner sep=0.75pt]  [font=\small] [align=left] {a)};
\draw (6,235.47) node [anchor=north west][inner sep=0.75pt]  [font=\small,rotate=-270] [align=left] {Transition Energy Variation (MHz)};
\draw (24.38,238.26) node [anchor=north west][inner sep=0.75pt]  [font=\small] [align=left] {b)};
\draw (318.67,32.21) node [anchor=north west][inner sep=0.75pt]  [font=\small] [align=left] {c)};
\draw (417.19,424) node [anchor=north west][inner sep=0.75pt]  [font=\small] [align=left] {Spectral Resolution $\displaystyle \sigma $ (MHz) };
\draw (322.05,300.05) node [anchor=north west][inner sep=0.75pt]  [font=\small,rotate=-270.08] [align=left] {Mean fit deviation (MHz) };
\draw (433.85,13) node [anchor=north west][inner sep=0.75pt]  [font=\small] [align=left] {$\displaystyle T_{2}$ Coherence Time (ns)};
\draw (108.47,444.57) node [anchor=north west][inner sep=0.75pt]  [font=\small] [align=left] {$\displaystyle Z-$field Strength $\displaystyle \epsilon/2\pi $ (GHz)};
\draw (441.69,362.5) node [anchor=north west][inner sep=0.75pt]  [font=\small,rotate=-270] [align=left] {$\displaystyle \sigma _{C}$ (MHz)};
\draw (502.19,373) node [anchor=north west][inner sep=0.75pt]  [font=\small] [align=left] {Number of Spins};
\draw (568.47,56.8) node [anchor=north west][inner sep=0.75pt]  [font=\small,color={rgb, 255:red, 255; green, 0; blue, 31 }  ,opacity=1 ]  {$\sigma _{C}( n=4)$};
\draw (8,449.47) node [anchor=north west][inner sep=0.75pt]  [font=\small,rotate=-270] [align=left] {Transition Energy Variation (MHz)};
\draw (105.83,325.37) node [anchor=north west][inner sep=0.75pt]   [align=left] {Hamiltonian Model Fit};
\draw (141.33,343.24) node [anchor=north west][inner sep=0.75pt]  [font=\small] [align=left] {4-local Model};
\draw (141.17,360.79) node [anchor=north west][inner sep=0.75pt]  [font=\small] [align=left] {3-local Model};
\draw (132.17,381.29) node [anchor=north west][inner sep=0.75pt]  [font=\small] [align=left] {Original Data Point};
\draw (384.52,94.33) node [anchor=north west][inner sep=0.75pt]  [font=\footnotesize] [align=left] {3-local Fit Deviation};
\draw (384.52,76) node [anchor=north west][inner sep=0.75pt]  [font=\footnotesize] [align=left] {4-local Fit Deviation};
\draw (373.85,59) node [anchor=north west][inner sep=0.75pt]  [font=\small] [align=left] {Hamiltonian Model Fit};
\draw (132.83,401.29) node [anchor=north west][inner sep=0.75pt]  [font=\small] [align=left] {Shifted Data Point};

\end{tikzpicture}

    \caption{Robustness of the detection method against emulated experimental error. (a) Transition energy variation as a function of $Z$-field for a specific field configuration before (red dots) and after (green dots) addition of a 5\,MHz error. The 3-local (4-local) fit to the noisy data is shown in orange (blue), and the fit deviation is the mean distance between fit and the noiseless data points. (b) Same as (a) but with an emulated error of 20\,MHz. (c) Comparison of the mean fit deviation between the 3- and 4-local models as a function of the spectral resolution $\sigma$. The critical noise amplitude, which is given by the intersection point $\sigma_C$, is highlighted with the red dashed line. Inset: Critical noise amplitude $\sigma_C$ versus number of spins $n$. We observe that $\sigma_C$ decreases exponentially with $n$ for a sufficiently large number of spins $n>2$.}
\label{fig:intersection}
\end{figure*}
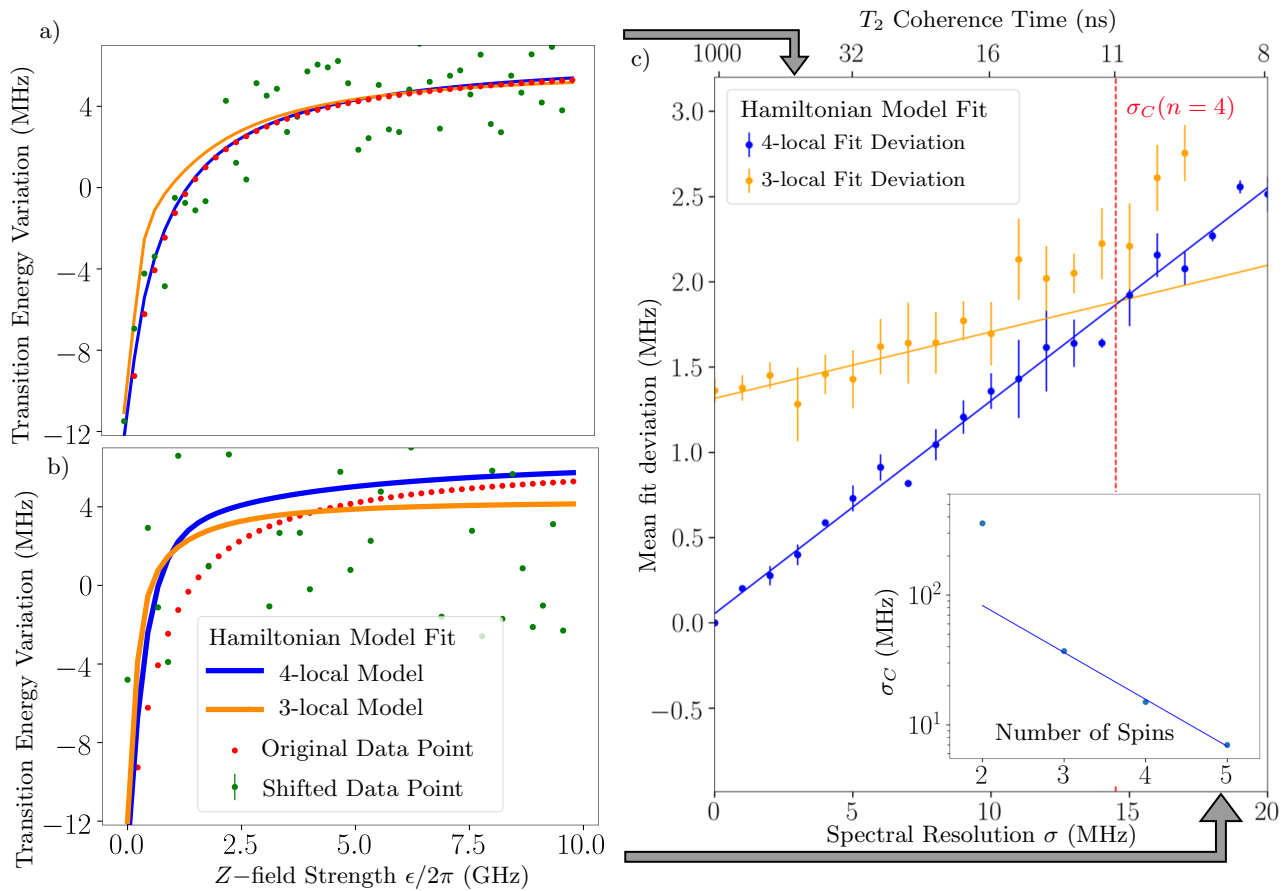

\textit{Spectral resolution threshold}.---We analyze the robustness of the proposed detection method against emulated experimental errors.
Specifically, we introduce noise to the generated data and analyze how both the $n$-local and $(n-1)$-local Hamiltonian models perform.
As we increase the noise amplitude $\sigma$, we expect the quality of the $n$-local model to worsen, until the quality of both models is comparable.
In this limit, we can no longer characterize whether the system exhibits $n$-local coupling or not.

To emulate noise with amplitude $\sigma$, we randomly shift each transition energy data point by an error $\xi \sim \mathcal{N}(0,\sigma^2)$ drawn from a normal distribution with mean $0$ and variance $\sigma^2$.
In this way, we emulate the spectral linewidth of a spin transition with coherence time $T_2 = 1 / (2\pi\sigma)$, which creates an uncertainty in the transition frequency measurement.
We then analyze the quality measure of both the $n$-local and $(n-1)$-local model as we increase $\sigma$.
We identify a critical noise amplitude $\sigma_C$ at which the quality of both models is comparable and we can no longer distinguish the Hamiltonian locality.

\figref{fig:intersection}(a-b) show the transition energy variation versus flux for a 4-spin system and two different error amplitudes.
The fit is performed simultaneously on all 15 field configurations but only one configuration is shown for clarity.
As expected, the $4$-local model mean deviation is small for small $\sigma$, while the $3$-local model mean deviation is relatively large, clearly distinguishing the two models.
When we increase $\sigma$, however, both Hamiltonian fits have large deviations and we cannot distinguish which model better describes the data.

We analyze this behaviour more precisely as a function of $\sigma$ in \figref{fig:intersection}(c): We increase $\sigma$ and again compare the mean fit deviation between the two models.
The critical noise amplitude $\sigma_C$ is defined by the intersection point of a pair of linear fits in \figref{fig:intersection}(c), in which the blue line fits to the $4$-local model data, and the orange line to the first five points of the $3$-local model data.
We highlight that at the intersection point $\sigma_C$ we can no longer distinguish between the models.
For four spins, the intersection occurs at around $\sigma_C(n=4) = 15\,\text{MHz}$, which corresponds to $T_2 = 11\,\text{ns}$.
This time is substantially lower than that of state-of-the-art superconducting qubit devices, which have coherence times in the microsecond regime \cite{PhysRevApplied.8.014004,Kjaergaard_2020}.

The extension of the error analysis for other system sizes is shown in the inset of \figref{fig:intersection}(c), and we analyze how the critical noise amplitude varies as a function of the number of spins $n$.
We find that $\sigma_C(n)$ decreases exponentially as a function of $n$, suggesting the need for exponentially more precise experiments and measurements with increasing system size.
A similar analysis is presented in the Supplementary Information regarding the effect of spurious terms on this detection technique.

\textit{Extension to larger system sizes}.---We now approach the detection problem from a more analytical perspective, examining how our proposed method generalizes asymptotically for large $n$.
To derive an asymptotic bound for $\sigma_C(n)$, we make the simplifying assumption that the system has no coupling terms beside the $n$-local term.
We assume that all $\delta_i$ parameters are identical to $\delta$ and that the $n$-local term $M$ is small compared to other permanent terms in the Hamiltonian.
Moreover, we consider a slightly different measure of fit quality: the mean deviation between the fit and the error-shifted data point, which allows easier theoretical treatment.

We show that $\sigma_C(n)$ varies asymptotically as $O\left(\frac{M}{2^n}\right)$:
We first give a brief argument for a lower bound of the quality measure of an $n$-local Hamiltonian fit as a function of $\sigma$.
Then we motivate an upper bound for the quality measure of an $(n-1)$-local Hamiltonian fit.
These bounds require the use of perturbation theory arguments, with the complete proof in the Supplementary Information. The main result is an exponentially decreasing bound for $\sigma_C$ of the form
\begin{align}
    \sigma_C &= O\left(\frac{M}{2^n}\langle \cos\theta\rangle^n\right),
\end{align}
where $\theta$ is the angle between the ground state and the $Z$-direction if the system were non-interacting. The exponentially decreasing bound shows us the limitations of the detection method. 
Our proof is straightforward to generalize to any spectroscopic detection method, as such a technique suffers from an exponentially decreasing error setback.
This is true while the system contains small coupling and permanent single spin $X$ terms, as this guarantees that $\langle \cos\theta\rangle<1$.
Our finding motivates the search for a more generalizable detection technique that can be compared to the spectroscopic method both for required control capabilities and asymptotic scaling.

\textit{Dynamic detection technique}.---In addition to the exponential bound for spectroscopic detection schemes, no experimental detection technique is expected to detect $n$-local coupling while maintaining spectroscopic errors greater than $O(M)$.
This implies that to detect an $n$-local coupling, we require our system to have a coherence time of at least $O(1/M)$.
We discuss a dynamic detection technique that achieves this asymptotically optimal bound without suffering from an exponentially decreasing error bound for larger $n$.
The approach is based on exploiting transitions which are prohibited without the presence of an $n$-local term.
Specifically, consider setting all $\epsilon_i$ parameters to zero, such that the permanent $\delta_i$ parameters are dominant and the system eigenstates are approximate $X$ eigenstates. The ground state then is $|\alpha \rangle = |-\rangle^{\otimes^n}$ and the most-excited state is $|\beta\rangle = |+\rangle^{\otimes^n}$. 

The coupler is pulsed periodically, and the $n$-local coupling term varies as $\delta H = M\cos\omega t \, Z_1\otimes Z_2\ldots\otimes Z_n$.
From first-order time-dependent perturbation theory we expect that there is a transition probability between the ground and the most excited state proportional to the matrix element $\langle\alpha|\delta H|\beta\rangle$.
We pick the pulsing frequency $\omega = \omega_{\alpha\beta}$ to maximize the transition probability, which then grows as $M^2t^2$ for small times.
To measure this effect, we assume that we can measure state components with some $n$-independent constant precision $\xi$.
To detect the state component we must wait a time $t_0$ such that $M^2t_0^2 = \xi\Rightarrow t_0 = O(\frac{1}{M})$.
This indicates that we require a $T_2$ coherence time that is at least as long, and thus we need a spectral resolution of $O(M)$. This dynamic detection technique is asymptotically optimal.

For the approach to be a valid detection technique, the lower-order coupling terms cannot contribute to the transition probability to first order.
Using the orthogonality of the eigenstates, we find that lower-locality terms give a zero matrix element.
Therefore, to first order in the strength of the $n$-local term $M$, we expect a non-zero transition probability only if the system has an $n$-local Hamiltonian.

The dynamic method generalizes better than the spectroscopic approach, since transition probabilities only need to be measured to a constant degree of precision, whereas the spectroscopic approach requires exponentially decreasing spectral resolution.
Another benefit is that the approach does not involve numerically expensive matrix diagonalization and fitting routines.
Nevertheless, a drawback is that the coupler must be pulsed, which is a relatively strong assumption given that existing multi-spin coupler proposals target annealing architectures with typically limited coherence times \cite{Schondorf_2019,melanson2019tunable,menke2019automated}.

We analyze whether the method is still viable in the presence of dephasing and energy decay, using the Lindblad Master equation for the system density matrix $\rho$:
\begin{equation}
\begin{split}
    \dot{\rho}(t) = &-i[H(t),\rho(t)] \\
     &+\sum_i \frac{1}{2}\big(2C_i\rho(t) C_i^{\dag} - \rho(t) C_i^{\dag}C_i - C_i^{\dag}C_i \rho(t)\big).
\end{split}
\end{equation}
We consider one collapse operator $C_i$ causing the decay of single spins: $C_i = \sqrt{\gamma_i}|-\rangle_i\langle +|_i$.
A second collapse operator causes dephasing: $C_i = \sqrt{\gamma_i/2} X_i$.
The dephasing rate $\gamma_i$ is chosen as $\gamma_i=1/T_2$, with energy decay and dephasing having equal rates.

\begin{figure}[t]
    \centering
\begin{tikzpicture}[x=0.75pt,y=0.75pt,yscale=-1,xscale=1]

\draw (172.5,116.06) node  {\includegraphics[width=222.75pt,height=152.66pt]{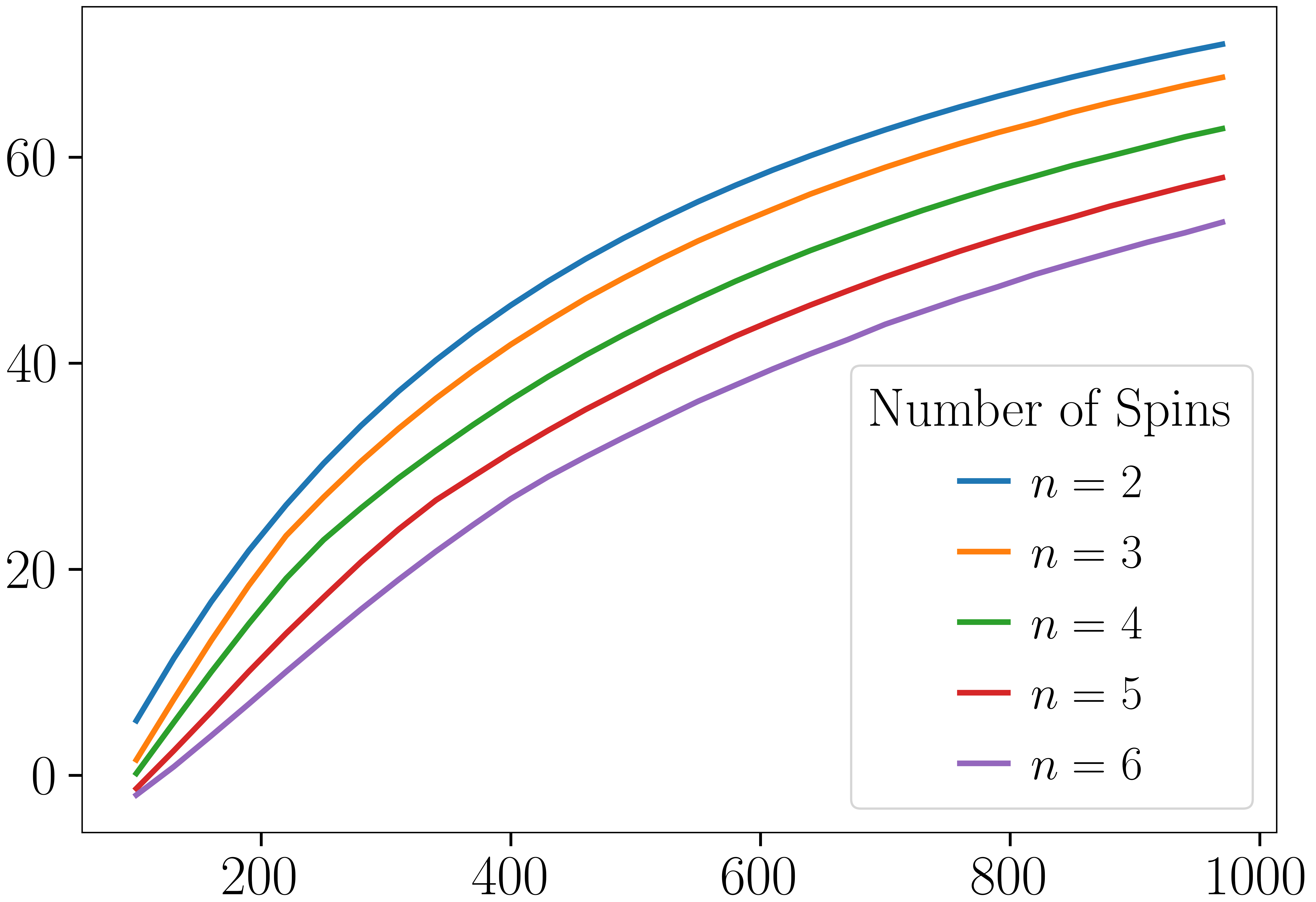}};
\draw  [color={rgb, 255:red, 255; green, 255; blue, 255 }  ,draw opacity=1 ][fill={rgb, 255:red, 255; green, 255; blue, 255 }  ,fill opacity=1 ] (256,114) -- (291,114) -- (291,193.29) -- (256,193.29) -- cycle ;
\draw  [color={rgb, 255:red, 255; green, 255; blue, 255 }  ,draw opacity=1 ][fill={rgb, 255:red, 255; green, 255; blue, 255 }  ,fill opacity=1 ] (219,99.29) -- (304,99.29) -- (304,116.29) -- (219,116.29) -- cycle ;

\draw (218,100) node [anchor=north west][inner sep=0.75pt]  [font=\footnotesize] [align=left] {Number of Spins};
\draw (260,115) node [anchor=north west][inner sep=0.75pt]  [font=\normalsize] [align=left] {$\displaystyle n=2$};
\draw (260.13,131.68) node [anchor=north west][inner sep=0.75pt]  [font=\normalsize] [align=left] {$\displaystyle n=3$};
\draw (259.1,146.85) node [anchor=north west][inner sep=0.75pt]  [font=\normalsize] [align=left] {$\displaystyle n=4$};
\draw (259.33,161.99) node [anchor=north west][inner sep=0.75pt]  [font=\normalsize] [align=left] {$\displaystyle n=5$};
\draw (259.5,178.38) node [anchor=north west][inner sep=0.75pt]  [font=\normalsize] [align=left] {$\displaystyle n=6$};
\draw (111.8,220) node [anchor=north west][inner sep=0.75pt]  [font=\normalsize] [align=left] {Coherence Time $\displaystyle T_{2}$ (ns)};
\draw (5,166.99) node [anchor=north west][inner sep=0.75pt]  [font=\normalsize,rotate=-270] [align=left] {State Contrast (\%)};

\end{tikzpicture}

    \caption{Plot of the state contrast as a function of the number of spins $n$ and coherence time $T_2$. For large enough coherence times $T_2$, the state contrast is high, which implies the existence of the desired $n$-local term.}
    \label{fig:coherence}
\end{figure}
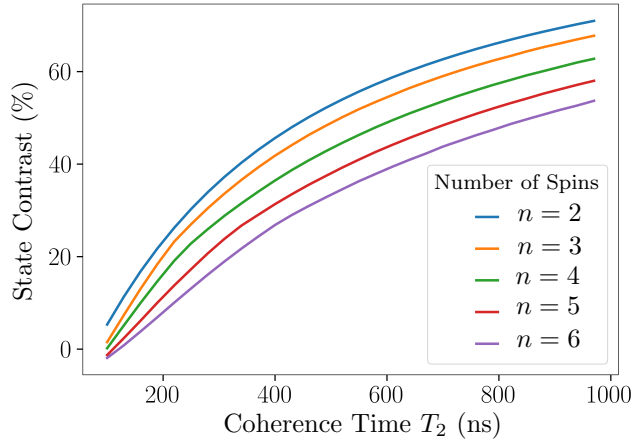

We analyze the method performance by analyzing the state contrast of the system during a period of $1000 \ \text{ns}$.
We define the state contrast as the difference between the maximum probability of measuring the $|+\rangle^{\otimes^n}$ state minus the maximum probability of measuring any other eigenstate during the time window, excluding $|-\rangle^{\otimes^n}$.
This is a natural measure of success, as the transition to the most excited state is predominantly due to the $n$-local term, and a sufficiently high state contrast would imply the existence of the desired term.

The time evolution of the system is simulated numerically with QuTiP \cite{Johansson_2013} and the result is shown in \figref{fig:coherence}.
We show the state contrast dependence on the number of spins $n$ and coherence time $T_2$.
For short coherence times, the state contrast is small, implying a sufficiently large coherence time is required. Furthermore, the state contrast decreases as $n$ is increased, which is expected as there are more decay pathways from the highest to lower-energy states.


\textit{Conclusion}.---We presented a spectroscopic detection technique capable of characterizing $n$-local spin interactions.
The robustness of the method against experimental errors and spurious terms was analyzed both numerically and analytically.
We found that this approach is expected to perform well for systems of up to five spins.
For larger systems, we contrasted the method with an alternative technique that relies on dynamical control of the coupler and is more extensible asymptotically.
For practical multi-spin coupler designs \cite{Schondorf_2019,melanson2019tunable,menke2019automated}, our methods are applicable for experimentally feasible error amplitudes, and thus can be used to characterize the system Hamiltonian and detect its locality.

We acknowledge Thiago Bergamaschi and Jeffrey A. Grover for valuable discussions.
This research was funded in part by the Office of the Director of National Intelligence (ODNI), Intelligence Advanced Research Projects Activity (IARPA) under Air Force Contract No. FA8702-15-D-0001. The views and conclusions contained herein are those of the authors and should not be interpreted as necessarily representing the official policies or endorsements, either expressed or implied, of the ODNI, IARPA, or the U.S. Government.


\bibliographystyle{apsrev4-1}
\bibliography{main}

\end{document}


\preprint{APS/123-QED}

\title{Supplementary Material: \\ A spectroscopic method for distinguishing multi-spin interactions \\ from lower-order effects}

\author{Thomas R. Bergamaschi}
\email{thomasbe@mit.edu}
\affiliation{\mitphysics}

\author{Tim Menke}
\email{timmenke@mit.edu}
\affiliation{\mitphysics}
\affiliation{\rle}
\affiliation{\harvardphysics}

\author{William P. Banner}
\affiliation{\miteecs}

\author{Agustin Di Paolo}
\affiliation{\rle}

\author{Steven J. Weber}
\affiliation{\mitll}

\author{Cyrus F. Hirjibehedin}
\affiliation{\mitll}

\author{Andrew J. Kerman}
\affiliation{\mitll}

\author{William D. Oliver}
\email{william.oliver@mit.edu}
\affiliation{\mitphysics}
\affiliation{\rle}
\affiliation{\miteecs}
\affiliation{\mitll}

\date{\today}

\maketitle

\section{Sensitivity to spurious terms}

As another measure of robustness, we analyze how the detection method performs in the presence of unwanted spurious fields and couplings.
If the spurious terms are large, they could suppress the influence of the $n$-local term on the transition energy spectrum, thus interfering with its detection. Therefore, it is important that the detection technique is capable of withstanding even large spurious terms.

We study the spurious term dependence similarly to the error analysis.
Specifically, for each coupling parameter $J$ in the original Hamiltonian presented in Eq.~(1) in the main text, we randomly shift it by a spurious coupling $\delta J$ when the coupler is turned on.
We choose $\delta J \sim U(0,\eta M)$, such that the spurious term is sampled from a uniform distribution between $0$ and $\eta M$, where $\eta$ is labeled the relative spurious term amplitude and $M$ is the $n$-local coupling term.
If the system had no experimental error, the $n$-local Hamiltonian fit would always be perfect independent of the spurious term.
In order to approximate an experimental realization of the system, however, we include an error of $\sigma =5$ \,MHz.

After applying all the $2(2^n-3)$ spurious terms $\delta J$ to the Hamiltonian and generating the data including noise, we again perform an $n$-local and an $(n-1)$-local Hamiltonian model fit.
We analyze the measure of quality of both models, examining whether we can always distinguish an $n$-local Hamiltonian from an $(n-1)$-local Hamiltonian.
For concreteness, we focus on a four-spin system and investigate the detection of a $4$-local interaction.

\begin{figure}[ht]
    \centering
    \begin{tikzpicture}[x=0.75pt,y=0.75pt,yscale=-1,xscale=1]

\draw (217.92,133.35) node  {\includegraphics[width=222.38pt,height=184.53pt]{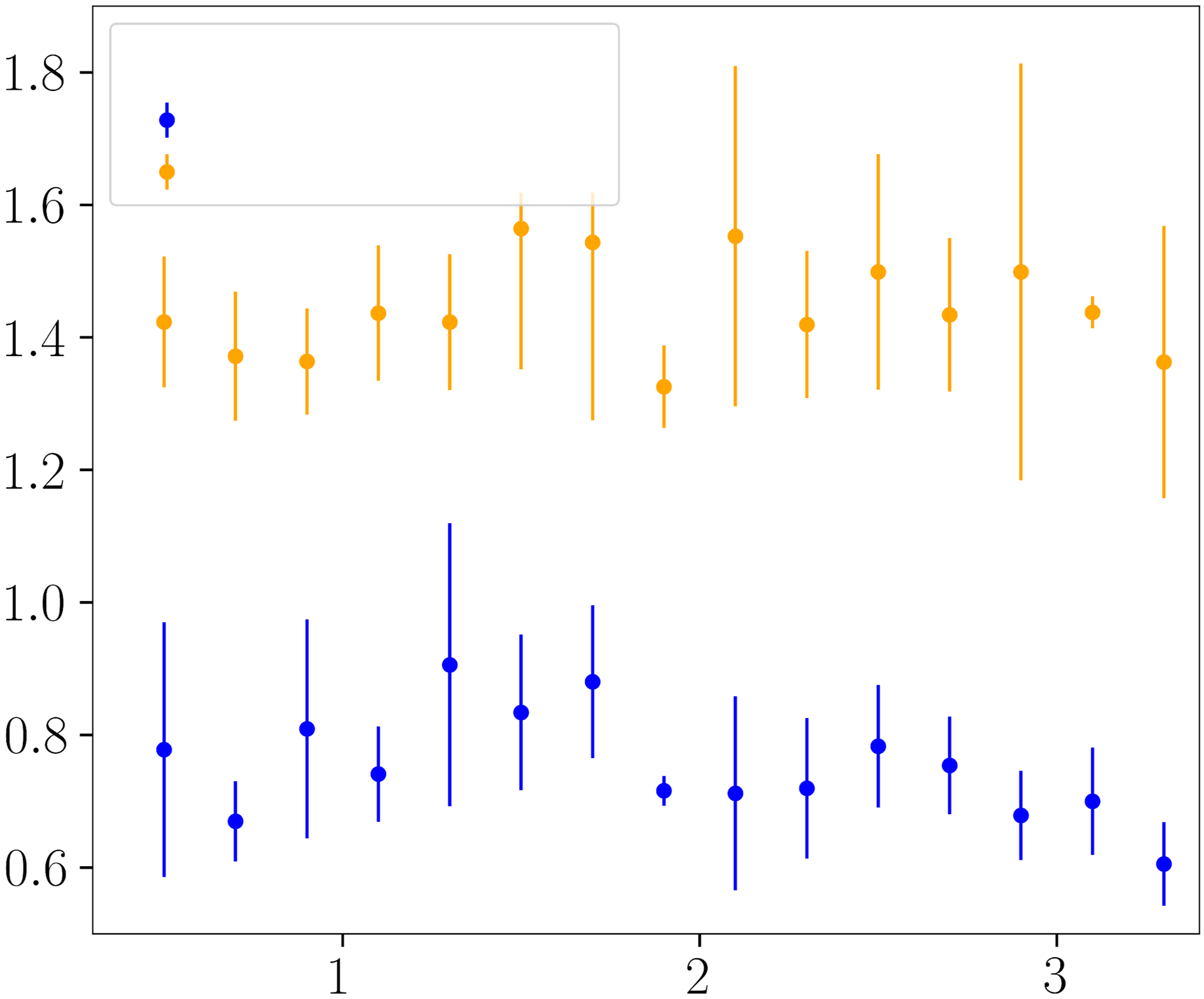}};

\draw (50.33,195) node [anchor=north west][inner sep=0.75pt]  [rotate=-270] [align=left] {Mean Fit Deviation (MHz)};
\draw (135.6,256.47) node [anchor=north west][inner sep=0.75pt]   [align=left] {Relative Spurious Term Amplitude};
\draw (99.63,18.35) node [anchor=north west][inner sep=0.75pt]  [font=\footnotesize] [align=left] {Hamiltonian Model Fit};
\draw (128.27,31.35) node [anchor=north west][inner sep=0.75pt]  [font=\footnotesize] [align=left] {4-local Fit};
\draw (128.27,46.02) node [anchor=north west][inner sep=0.75pt]  [font=\footnotesize] [align=left] {3-local Fit};

\end{tikzpicture}

    \caption{(a) Mean fit deviation for the 3-local (orange) and 4-local (blue) models as a function of the relative spurious term amplitude $\eta$. There is a significant separation between both curves, which indicates that even large spurious terms do not pose a problem for the detection method.}
    \label{fig:spurious}
\end{figure}

Interestingly, we observe from the data in \figref{fig:spurious} that the spurious term amplitude does not present difficulties in detecting the desired $n$-local Hamiltonian over an $(n-1)$-local Hamiltonian, as there is always a significant distance between the two measures of quality. Therefore, the spurious term amplitude does not complicate the detection of the $n$-local term, such that the main impediment of this detection technique is experimental errors in the spectroscopic measurements.

\section{Proof of the Asymptotic error cutoff}

Here we present the proof of the asymptotic error cutoff $\sigma_C$ as a function of the number of spins $n$.
To do so, we first note that the $n$-local Hamiltonian model fit on noisy $n$-local data will necessarily have a measure of quality that scales at least linearly with $\sigma_C$. In order to upper bound the measure of quality of an $(n-1)$-local Hamiltonian fit on noisy $n$-local Hamiltonian data, note that any such $(n-1)$-local fit on this data will perform at least as well as the $(n-1)$-local fit that finds all spurious parameters correctly but has no $n$-local coupling term.
Under the assumption of small coupling terms, the system's eigenstates are approximate non-interacting states, such that the ground state is $\otimes_{i=1}^n |g_i\rangle$.
One of the first excited states is $|f_1\rangle \otimes_{i=2}^n |g_i\rangle$, where $|g_i\rangle = -\sin\frac{\theta}{2}|0\rangle + \cos\frac{\theta}{2}|1\rangle$ and $|f_i\rangle = \cos\frac{\theta}{2}|0\rangle + \sin\frac{\theta}{2}|1\rangle$.
The angle $\theta_i$ is given by $\cos\theta_i = \frac{\epsilon_i}{\sqrt{\delta^2+\epsilon_i^2}}$.

To compute how the desired $n$-local Hamiltonian compares to the $(n-1)$-local Hamiltonian with all spurious parameters correct, we apply perturbation theory. Specifically, the difference between each curve is easily computable to first order in $M$, and we find that it is $2M(-1)^n\prod_i \cos\theta_i$. To this order, this is simply the value of the deviation between this Hamiltonian's data and the original $n$-local Hamiltonian data. Observe that this is always $0$ if for some $i$ $\cos\theta_i = 0\Rightarrow \epsilon_i = 0$, such that we only expect a difference if we are in the configuration that sweeps all $\epsilon$ parameters together with $\epsilon_i = \epsilon$. Therefore, the mean deviation is $2\frac{M}{2^{n}-1}\langle \prod_i \cos\theta_i\rangle = 2\frac{M}{2^{n}-1} \langle \cos\theta\rangle^n = O(\frac{M}{2^n}\langle \cos\theta\rangle^n)$, where we use $\langle \cdot \rangle$ to indicate the average value over all $\epsilon$. In this case $\langle \cos\theta\rangle = \langle \frac{\epsilon}{\sqrt{\epsilon^2+\delta^2}}\rangle$, where we sweep $\epsilon$ from $0$ to $\epsilon_{\max}$, such that $\langle \cos\theta\rangle = \frac{\sqrt{\delta^2+\epsilon_{\max}^2}-\delta}{\epsilon_{\max}}<1$.

As the optimal $(n-1)$-local Hamiltonian fit performs at least as well as this constructed Hamiltonian, this implies that the measure of quality of the $(n-1)$-local fit is at most $O(\frac{M}{2^n}\langle \cos\theta\rangle^n)$. Therefore, we expect an error amplitude cutoff at $\sigma_C = O(\frac{M}{2^n}\langle \cos\theta\rangle^n)\leq O(\frac{M}{2^n})$, an exponentially decreasing bound as claimed.


\section{Second order transition probabilities}

In this section, we consider that there are non-zero lower-order coupling terms and examine the effect of these on the dynamic detection method. We consider a system Hamiltonian with large $X$ field terms, such that the eigenstates are approximate $X$-eigenstates. Specifically, we label the lowest and highest excited state $|i\rangle = |-\rangle^{\otimes^n}$ and $|f\rangle = |+\rangle^{\otimes^n}$ respectively.

In addition, we assume that the spurious terms caused by the coupler oscillate with the same frequency as the $n$-coupling term. Specifically, we assume a harmonic perturbation of the following form, which is slowly turned on at $t=-\infty$:
\begin{equation}
\label{eq:perturbation}
    V(t) = e^{\epsilon t}\cos(\omega t)\sum_{Q\subseteq S} \delta J_{Z}^{(Q)} \Motimes_{s\in Q} Z_{s} + \delta J_{X}^{(Q)} \Motimes_{s\in Q} X_{s}.
\end{equation}
We later take the limit $\epsilon\rightarrow 0$.

Again, we set $\delta J_X^{(S)}=0$ and for convenience set $\delta J_Z^{(S)}=M$.
Using the same argument as in the main text, the lower locality terms in $V(t)$ do not contribute to first order in the transition probability.
By second order perturbation theory for the time dependent perturbation $V(t)$, the transition probability can be written as a power series:

\begin{equation}
    P_{|i\rangle\rightarrow |f\rangle}(t) = |c(t)|^2 \text{ with } c(t) = c^{(1)}(t) + c^{(2)}(t)\ldots, \text{ where } c^{(n)}\sim O(V^n).
\end{equation}

For $c^{(1)}(t)$ we consider first order perturbation theory as in the main text, and we find $c^{(1)}(t) = Mt$.
For the second order correction $c^{(2)}(t)$ we have
\begin{equation}
\label{eq:pert theory}
    c^{(2)}(t) = -\sum_{m} \int_{-\infty}^t dt'\int_{-\infty}^{t'} dt'' e^{i \omega_{fm} t' + i \omega_{mi}t''}V_{fm}(t')V_{mi}(t''),
\end{equation}
where $\omega_{fm} = \omega_f-\omega_m$ is the energy difference between the states $f$ and $m$, and $V_{fm}(t) = \langle f|V(t)|m\rangle$.
With this in mind, we compute $V_{fm}(t)$: The key observation is to consider expanding the perturbation $V(t)$ in the sum format.
Exploiting orthogonality, the only terms that survive in this term are the terms in $V(t)$ that when acting on $|m\rangle$ produce a term that includes $|f\rangle$.
For the $f=m$ case, this is the term $\sum_{Q\subseteq S} \delta J_X^{(Q)}$.
For the $f\neq m$ case, we note that the only term that can connect $\ket{m}$ and $\ket{f}$ is the $Z$-coupling term that acts on the appropriate subset of spins $Q$ such that $\Motimes_{s\in Q} Z_{s} |m\rangle \propto |f\rangle$.
As $|m\rangle$ and $|f\rangle$ are $X$ eigenstates, we require that $Q = \{\bar{m}\}$, where by $\{\bar{m}\}$ we denote the set of spins in $m$ which are in the $|-\rangle$ state.
With this, we find that $V_{fm}(t) = \cos \omega t \, \bigg(\delta_{fm} \sum_{Q\subseteq S} \delta J_X^{(Q)} + \delta J_Z^{(\bar{m})}\bigg) = \tilde{J}_{fm}\cos\omega t $.
Analogously for $V_{im}(t) = \cos \omega t \, \bigg(\delta_{im} \sum_{Q\subseteq S} \delta J_X^{(Q)} + \delta J_Z^{(m)}\bigg) = \tilde{J}_{im}\cos\omega t $, where we denote the set $\{m\}$ as the set of spins in $m$ in the $|+\rangle$ state.

Therefore, we find that a single component inside the sum over $m$ in \eqref{eq:pert theory} is:
\begin{equation}\label{eq:pert}
    \int_{-\infty}^t dt'\int_{-\infty}^{t'} dt'' e^{i \omega_{fm} t' + i \omega_{mi}t''}V_{fm}(t')V_{mi}(t'') = \int_{-\infty}^t dt'\int_{-\infty}^{t'} dt'' e^{i \omega_{fm} t' + i \omega_{mi}t''} \tilde{J}_{im}\tilde{J}_{fm} \cos(\omega t')\cos(\omega t'')
\end{equation}

Using that we are setting $\omega = \omega_fm$, this reduces the integral in \eqref{eq:pert} to $-\frac{e^{-i\omega t}}{\omega \omega_{fm}}\tilde{J}_{im}\tilde{J}_{fm}$.

The overall coefficient $c^{(2)}(t)$ computes to
\begin{equation*}
    c^{(2)}(t) = - \frac{e^{-i\omega t}}{\delta \omega}\sum_m \frac{\tilde{J}_{im}\tilde{J}_{fm}}{\omega_{fm}} \sim  - \frac{e^{-i\omega t}}{\delta \omega}\sum_m \bigg(\delta_{im} \sum_{Q\subseteq S} \delta J_X^{(Q)} + \delta J_Z^{(m)}\bigg)\bigg(\delta_{fm} \sum_{Q\subseteq S} \delta J_X^{(Q)} + \delta J_Z^{(\bar{m})}\bigg)=
\end{equation*}
\begin{equation}
    \sim - \frac{e^{-i\omega t}}{\delta \omega}\bigg(\sum_m \delta J_Z^{(m)}\delta J_Z^{(\bar{m})} + 2\delta J_Z^{(f)}\sum_{Q\subset S} \delta J_X^{(Q)}\bigg).
\end{equation}
Now we note that $\delta J_Z^{(f)} = M$ and for simplicity assume all $\delta$ parameters in the original Hamiltonian are equal such that $\omega = 2n\delta$.
We highlight the contribution in the above formula that is not due to the $n$-local coupling.
This is the contribution from the term $\sum_m \delta J_Z^{(m)}\delta J_Z^{(\bar{m})}$, which to be non-zero requires that there exists at least one combination of $2$ spurious terms caused by the coupler: one of locality $k$ and the other of locality $n-k$, for any positive $k$.
It is important to highlight that if such combinations can be avoided, for example by the coupler design, then to second order in perturbation theory the transition probability will be solely due to the $n$-coupling term.
Similar arguments extend to higher orders in perturbation theory.

Thus, to second order in perturbation theory and assuming all coupling parameters have equal size $\delta J$, the contribution to $c^{(2)}(t)$ due to the spurious terms is $2^n \frac{\delta J^2e^{-i\omega t}}{2n\delta^2}$.
For the transition probability to be predominantly caused by the $n$-local term, we find that $M/\delta$ must be considerably larger than the spurious term effect.
Ignoring the complex exponentials, this implies that $M$ must be larger than $2^{n-1} \frac{\delta J^2}{n\delta}$.
We highlight that as most spurious terms have size $\delta J = O(M)$, the dynamic detection technique requires $\delta \gg \frac{2^{n-1}}{n}M$.
This bound is more achievable in practice for large $n$ than the exponential bound on the spectral resolution of the static technique.
That is because $M$ is expected to be small in the first coupler prototypes.

\bibliographystyle{apsrev4-1}
\bibliography{main}